\documentstyle[12pt]{article}   

\large
\newcommand{\tiN}{\raisebox{-6.5pt}{$\displaystyle
\stackrel{\displaystyle N}{\sim}$}}
\newcommand{\tiM}{\raisebox{-6.5pt}{$\displaystyle
\stackrel{\displaystyle M}{\sim}$}}
\newcommand{\beq}{\begin{equation}}
\newcommand{\eeq}{\end{equation}}

\makeatletter
\@addtoreset{equation}{section}
\makeatother

\title{Reduced models for quantum gravity}
\author{T. Thiemann\thanks{thiemann@phys.psu.edu} \\
       {\small Institut f\"ur Theoretische Physik}\\
       {\small RWTH Aachen}\\
       {\small Sommerfeldstrasse, D-52056 Aachen, Germany}   
\thanks{present address : Center for Gravitational Physics and Geometry,
        The Pennsylvania State University, 
        University Park, PA 16802, USA}}
\date{}

\begin{document}

\maketitle                     

\section{Introduction}

As outlined in various other lectures given at this meeting, it seems that a 
quantum theory of
gravity can only be constructed in a non perturbative manner (compare, in 
particular, Ashtekar's lectures).

Because of that, no calculations, as for example cross sections, decay rates
and so on, can be done for the full theory of 3+1
gravity unless one has the full solution space of the quantum constraints
and, derived from that, the physical Hilbert space. For a non-gravitational
quantum field theory that can be attacked via a perturbative
approach one {\em can} make quantitative predictions and even make
estimates of the error due to higher order corrections while for quantum
gravity 'one would have to consider {\em all orders}'. So for 
the former theories
one {\em does} have a very good idea of how the exact quantum theory should 
look like and this is important because intuition gives rise to new lines 
of attack.\\
Hence unfortunately, for the quantum theory of gravity, we lack this general 
picture
of how the exact theory 
should look like completely.

The only way that might help to uncover some of the secrets of how to solve 
the technical and/or conceptual problems of quantum gravity seems to be to
study model systems, ideally those that can be solved exactly.

Of course, the lessons that models teach us might be totally misleading and
extreme care is due when transferring results from the model to the physical
theory of full quantum gravity.

This is the point of view that we adopt in the sequel : \\
The models that we are going to discuss capture some of the technical and
conceptual problems of gravity and we will pin these down. We {\em will}
attempt
at drawing some conclusions from the solutions we found but we {\em stress}
the limitations that arise from the various special features of the models
we choose and which are not shared by the full theory of quantum gravity.

From the various, completely solvable, models that have been discussed in
the literature we choose those that we consider as most suitable for our
pedagogical reasons, namely 2+1 gravity and the spherically symmetric
model.

The former model arises from a dimensional, the latter from a
Killing reduction of full 3+1 gravity. While 2+1 gravity is usually
treated in terms of closed topologies without boundary of the initial data
hypersurface, the topology for the spherically symmetric system is chosen to
be asymptotically flat. Finally, 2+1 gravity is more naturally quantized
using the loop representation while spherically symmetric gravity is
easier to quantize via the self-dual representation.

Accordingly, both types of reductions, both types of topologies and both
types of representations that are mainly
employed in the literature in the context of the new variables come into
practice.

It is true 
that both models turn out to have only a finite-dimensional reduced phase
space, hence, we are actually dealing only with a quantum mechanical
problem and this restricts the usefulness of the results found for the
models tremendously when transferring them to quantum gravity which
is a theory with an infinite number of physical degrees of freedom.
On the other hand, models with a finite number of degrees of freedom make the
analysis especially clear and 
since one of the major motivations of this meeting was that it
should serve as an introduction to canonical quantum gravity,
we regard it as important to demonstrate the usefulness of the formalism
that has been developed in various other lectures of this seminar
by means of applying it to models of quantum gravity which are as simple as 
possible, here
formulated in terms of Ashtekar's new variables.

We adopt
the abstract index notation and also the notation used by the the other
contributors to the seminar.

\section{2+1 gravity}

\subsection{Canonical formulation}

The model of 2+1 gravity arises from a dimensional reduction of the
Einstein action of full 3+1 gravity. The first order (Palatini) formulation
of n+1 gravity is given by
\beq S:=\frac{1}{2\kappa}\int_M \Omega_{IJ}\wedge\star(e^I\wedge e^J) \eeq
where the notation is as follows : M is the spacetime manifold 
(dim(M)$=n+1$), $\kappa$ is
the gravitational constant, $\Omega_{IJ}$ is the curvature 2-form of the
SO(1,n) principal connection $\omega_{IJ}$, $\star$ is the Hodge-duality 
operator
and $e^I$ are the (n+1)-bein fields (that is, an orthnormal cobasis field).
Internal indices I,J,K,.. run from 0 to n and are raised and lowered with
respect to the internal Minkowski metric $\eta_{IJ}$.\\
For the canonical formulation we have to do the n+1 split of the action
by choosing a foliation of M into space and time, that is, we assume that
M is topologically $\Sigma\times R$ where $\Sigma$ is an n-dimensional
spacelike hypersurface. From now on we will restrict ourselves to the case
$n=2$.\\
As in \cite{1}, we choose icoordinates ${x^0,x^1,x^2}$
and a foliation such that the label of the various hypersurfaces is given
by $t:=x^0$. Then we obtain from (2.1)
\beq S=\frac{1}{2\kappa}\int_R dt\int_\Sigma[\dot{A}_a^I E^a_I-[-\Lambda^I
{\cal G}_I+N_I C^I]] \; . \eeq
We now explain the notation :
The indices a,b,c,.. are 2-valued and have the meaning of tensor indices
with respect to the spatial slice. $A_a^I$ is simply the pull-back to the
spatial slice of the 3-dimensional connection $\omega^I$ and we
have exploited the fact that the defining and adjoint representation
of SO(1,2) are isomorphic such that we can work with the quantities
$\omega^I:=-\frac{1}{2}\epsilon^{IJK}
\omega_{JK}$. As is obvious from the action (2.1), in this polarization
\footnote{a polarization is, roughly, a subdivision of a choice of phase
space coordinates, into momenta and configuration space variables}
the SO(1,2) connection $A_a^I$ will play the role of a configuration space
variable.\\
Its conjugate momentum is evidently given by the 'electric fields'
$E^a_I:=\epsilon^{ab}e_b^I\mbox{ where }e_b^I$ is the pull-back 
of the SO(1,2) triad.
Geometrically, the electric fields are so(1,2)-valued
vector fields of density weight one due to the 2-dimensional 
(metric-independent) Levi-Civita
density $\epsilon^{ab}$.\\
The 'Hamiltonian' is a pure constraint. Two kinds of constraints arise.
The Gauss constraint
\beq {\cal G}_I:={\cal D}_a E^a_I:=\partial_a E^a_I+\epsilon_{IJ}\;^K A_a^J
E^a_K \eeq
is manifestly reminiscent of the Gauss constraint for 3+1 gravity while
the other 3 constraints
\beq C^I:=B^I=\frac{1}{2}\epsilon^{ab}F_{ab}^I \eeq
tell us that that the magnetic fields (equivalently the curvature $1/2
F_{ab}^I\mbox{ of }A_a^I$) vanish, that is, the constraint surface of the
phase space contains only flat connections.\\
It follows from general arguments that the constraints form a first class
subalgebra of the Poisson algebra of functions on the phase space since
they are either linear in or independent of the momenta.\\
Immediately, the question arises what these 'flatness' constraints have to
do with the constraints of 3+1 gravity (compare Giulini's lectures).
The answer is that, provided that the twice densitized inverted 2-metric
\beq E^a_I E^b_J \eta^{IJ} \eeq
is nondegenerate, then
we can recast these constraints in an 'Ashtekar-like' form :
\begin{eqnarray}
V_a & := &\epsilon_{ab}B^I E^b_I\mbox{ : vector constraint and} \\
C & := & \epsilon_I\;^{JK}B^I\epsilon_{ab} E^a_J E^b_K\mbox{ : scalar
constraint.}
\end{eqnarray}
We can interpret this as follows : \\
Choose the Lagrange multipliers 
$\Lambda^I:=-\omega_t^I,\; N_I:=-e_{tI}$ such that
\beq N_I=\frac{1}{2}(\tiN\epsilon_I\;^{JK}E^a_J E^b_K+N^a E^b_I)\epsilon_{ab}
\eeq
where $N:=\sqrt{\det(q)}\tiN,\; N^a$ are lapse and shift functions and the 
first 
term in (2.8) is
non-vanishing if and only if (2.5) is invertible. Then
$N_I C^I=N^a V_a+\tiN C$.\\
My personal point of view is that in order to test 3+1 gravity one should
actually start from the constraints (2.6) and (2.7) rather than from (2.4) and
allow for general, in particular non-flat, connections although the
induced 2-metric on the 2-dimensional slice is then in general singular.
This is because in 3+1 gravity the first task in the quantization programme
is the solution of the constraints and therefore any model should mirror
the algebraic form of the constraints as closely as possible. Further, one
quantizes 3+1 gravity in the connection representation and thus the state
functional will in general have support on non-flat connections.\\
There has already been done some work in that direction (\cite{2}) : the 
authors
of that paper show that 2 large classes of solutions to the vector and
scalar constraint in this 'degenerate' sector allow for a reformulation
of (2.4) such that one has the vector constraint and either 1) that
the magnetic field $B^I$ is null or 2) that the 2 internal vectors
$E^1_I, E^2_I$ are colinear.\\
However, since no closed solution to 2+1 gravity for the degenerate sector
is known yet, we consider it more appropriate for the present purpose 
to proceed with the non-degenerate (or Witten-) sector for which the
connection is constrained to be flat.\\
There is still another, quite important, difference between 2+1 gravity and
3+1 gravity : in the 2+1 case the connection is manifestly real whereas
in the 3+1 case it is genuinely complex, in general. Since the reality
conditions play a major role in the process of selecting an inner product
as outlined in earlier lectures (compare those of Giulini, Hajicek and
Rendall) we expect that the inner products of the two theories will not
be closely related to each other.\\
To complete the canonical formulation, we have to choose the topology of the 
initial data hypersurface $\Sigma$. In order to avoid technical issues
that have to do with the choice of function spaces to which the fields
belong (essentially, a choice of fall-off properties at infinity) we will
choose a closed topology without boundary as is common in the literature
(\cite{3}). The classification of these topologies is well-known. The 
characterizing parameter is
the genus g (number of handles) of $\Sigma$.

We will choose later the case of (g=1) (torus) since in this case the
quantum theory can be constructed in closed form. We will largely follow
(\cite{4}) in the sequel.

\subsection{The reduced phase space}

We will quantize the present model via the reduced phase space approach,
that is, we will determine the gauge invariant information of the phase
space before quantizing.\\

Up to a small degeneracy (\cite{5}) that arises for non-compact gauge
groups, the set of traces of the holonomy
\beq T^0_\alpha:=\mbox{tr}[h_\alpha[A](t)]
:=\mbox{tr}[{\cal P}\exp(\oint_\alpha A)] \eeq
is a good coordinate for the reduced configuration space with respect to the
gauge constraint. Here, $\alpha$ is any loop (i.e. an embedding of the
circle into $\Sigma$), $\cal P$ means path-ordering and the trace is taken
with respect to the 2-dimensional fundamental representation of SU(1,1)
(thereby exploiting that the Lie algebras su(1,1) and so(1,2) are
isomorphic). A suitable basis of su(1,1) is given by
\beq \tau_0:=-\frac{i}{2}\sigma_3,\;\tau_1:=\frac{1}{2}\sigma_1,\;
\tau_2:=\frac{1}{2}\sigma_2 \eeq
where $\sigma_I$ are the usual Pauli matrices (the index I is 0,1,2) with
respect to which the structure constants are given by $\epsilon_{IJ}\;^K$.
The holonomy itself depends on the starting point $\alpha(t)$ of the loop
(t ranges from 0 to 1 (to be identified) in our choice of parametrization)
however the $T^0$'s are independent of the starting point.
The reason why (2.9) is SU(1,1) invariant (even under large, rather than
infinitesimal SU(1,1) transformations) is that the holonomy is conjugated
by a gauge transformation U which drops out under the trace.\\
The $T^0_{\alpha}$ are even invariant under the gauge transformations
generated by the constraints $C^I$ because $C^I$ are independent of the
momenta so that the Poisson bracket\newline $\{T^0_\alpha,
\int_\Sigma d^2x N_I C^I\}$ vanishes. Accordingly, $T^0_\alpha$ is a Dirac
observable by definition.\\
Next, we need to capture gauge invariant information about the momenta. As
is suggested by the construction of the loop variables in the 3+1 case
(compare Br\"ugmann's lectures) we consider the 'smeared' version of the
so-called $T^1$ variables :
\beq T^1_{\alpha}:=\int_0^1 ds \dot{\alpha}^b(s) \epsilon_{ab}
\mbox{tr}[E^a(\alpha(s))h_\alpha[A](s)] \;. \eeq
Since $E^a=E^a_I\tau^I$ transforms according to the adjoint representation
of SU(1,1), these variables are manifestly gauge invariant. The crucial
step is to check its behaviour when taking the Poisson bracket with respect
to the flatness constraint. We find
\begin{eqnarray}
\{T^1_\alpha,\int_\Sigma d^2x N_I B^I\} & = & \int_0^1 ds \dot{\alpha}^a(s)
\mbox{tr}[\tau_I h_\alpha(s)]{\cal D}_a N^I(\alpha(s))\nonumber \\
& = & \lim_{t\to 0}\frac{1}{t}\{T^0_\alpha[A+t{\cal D}N]-T^0_\alpha[A]\}=0
\end{eqnarray}
because $T^0$ is gauge invariant and $A_a^I\to A_a^I+t{\cal D}_a N^I$ is an
infinitesimal gauge transformation. Hence the $T^1$'s are also Dirac
observables. Finally, again referring to \cite{5}, we learn that up to
a small degeneracy the $T^0,\;T^1$ capture all the information on the
reduced phase space.\\
On the constraint surface, both Dirac observables have support only on
flat connections. It follows that $T^1_\alpha$ only depends on the
homotopy class of the loop $\alpha$, denoted $[\alpha]$, when restricted to
the constraint surface. This can be seen as follows : deform the loop
$\alpha$ infinitesimally to get a new loop $\alpha'$. It then follows
from the Ambrose-Singer theorem (e.g. \cite{6}) that
$T^0_{\alpha'}-T_\alpha$ can be expanded in positive powers of coefficients
of the curvature of $A_a^I$ which vanishes on the constraint surface.
Therefore, $T^0_\alpha$ is (weakly) nontrivial only if $\alpha$ is
not contractible. The structure of the reduced phase space will therefore
be largely governed by the choice of genus which determines the dimension
of the homotopy group of $\Sigma$.\\
Choose a basepoint * in $\Sigma$ and consider all loops starting and
ending at *. The composition of loops equips this set of loops with the
structure of a semigroup (it is not a group since $\alpha\circ\alpha^{-1}
\not=*$, the trivial loop, and $\alpha^{-1}$ is the loop $\alpha$ traversed
in opposite direction). However, $h_\alpha h_{\alpha^{-1}}=1$. Thus, when
identifying loops $\alpha,\; \beta$ according to the rule $\alpha\approx\beta
\mbox{ iff } h_\alpha(h_\beta)^{-1}=1$
we get a group homomorphism
\beq \alpha\to h_\alpha[A]\mbox{ such that }h_{\alpha^{-1}}=(h_\alpha)^{-1}
\eeq
from the so constructed group of loops modulo $\approx$ (called the hoop
group) into SU(1,1) for any connection A.\\
Now, as derived above, the holonomy actually only depends on the homotopy
class of the hoop $\alpha$, that is $h_\alpha=h_{[\alpha]}$. Therefore,
we obtain a homomorphism from the homotopy group of hoops (called the set
of equitopic hoops in \cite{4}) into SU(1,1).\\
We now make use of the following fact (\cite{17}, Barrett's theorem) : there
is a bijection between (smooth, in Barret's topology) homomorphisms from the 
group of hoops into the 
gauge
group under consideration and (smooth, in the usual sense of smooth functions) 
connections of the associated 
principal fibre bundle up to gauge equivalence (that means that we have
only a one to one correspondence between homomorphisms and gauge equivalence
classes of connections).
Accordingly, we may think from now on of (the gauge equivalence class of) a 
given
connection
as given by the set of all possible   
smooth homomorphisms $h_\alpha$ from the fundamental group
$\pi_1(\Sigma)$ of the hypersurface into SU(1,1).\\
Up to now, the discussion was valid for arbitrary genus. We now specialize
to the torus. Exploiting that the homotopy group of the torus is abelian
(in terms of generators and relations, for genus g the homotopy group
consists of 2g generators and one relation), compare \cite{1}, we have that
for any {\em flat} connection A the holonomies of 2 hoops commute :
\beq h_\alpha h_\beta=h_{[\alpha]\circ[\beta]}=h_{[\beta]\circ[\alpha]}
=h_\beta h_\alpha \eeq
which is only possible if for any flat A all the $h_\alpha[A]$ lie in the same
abelian
subgroup of SU(1,1). We conclude that the homomorphism therefore must 
take the form $h_\alpha=\pm\exp(a(\alpha)t^I \tau_I)\mbox{ where }a(\alpha)$
is some real number depending on the loop $\alpha$ and $t^I$ is some constant
(loop-independent) internal vector. The 2 possible signs of the exponential
capture the fact that in SU(1,1) not every group element can be written as
the exponential of an element of the Lie algebra. This is due to the fact
that the group element $U=-1$ can be written in such a way only if the vector
$t^I$ is timelike : $-1=\exp(\tau_I t^I 2\pi)$.\\
Let us write down, for illustrative reasons, a connection in a particular 
gauge such that we get the
above homomorphism with the positive sign :\\
$A_a^I(x)=f_a(x) t^I$.
The flatness condition shows that f is a closed one form. It can therefore
be labelled by the cohomology class to which it belongs. The 2 generators
of the first cohomology group of $\Sigma$ are just the two anglar coordinates
$x^i,\;i=1,2$ of the torus such that up to a total differential
$f_a=a_i x^i_{,a}$ and $a_i$ are real numbers.\\
Explicitely we compute
\beq h_\alpha[A]=\exp(t^I \tau_I a_i\oint_\alpha dx^i) \eeq
which under a SU(1,1) transformation $U\; : \;\Sigma\to SU(1,1)$ becomes
\[ h_\alpha[A]\to U(*)h_\alpha[A]U^{-1}(*)=\exp(t^I U(*)\tau_I U^{-1}(*) a_i
\oint_\alpha dx^i)\; , \]
that is, the internal vector $t^I$ undergoes a constant (since U(*) only
depends on the basepoint *) gauge transformation thereby preserving only
its timelike, spacelike or null character.\\
Note that we could choose the function f in $A_a^I=f_{,a}t^I$ in such a way
along the loop $\alpha$ that it is smooth everywhere, vanishing at the 
parameter values 0 and $1/2$ and so that $t^I$ is a constant timelike vector
on the first half
of the loop and a constant spacelike vector on the second half of the
loop while $\int_0^{1/2}dt \dot{f}
(\alpha(t))=2\pi$. Then the holonomy of that connection is still of the 
general form $\pm\exp(a(\alpha)\tau_I t^I)$ and there are indeed connections 
that accomodate for both signs.\\   
Returning to the general case, in order to characterize our homomorphism
completely it is sufficient to give the image of the two generators 
$[\alpha_i]$
of the homotopy group $\pi_1(T^2)=Z\times Z$. This is given by either
$+\exp(a_i \tau_I t^I) \mbox{ or }-\exp(a_i \tau_I t^I)$ where $a_i$ are again
two real numbers.\\
Let us choose standard internal vectors with norm, respectively, $\pm 1,0$ in 
the spacelike, timelike or null sector respectively. In the non-spacelike case
these can also be taken to be future directed while in the spacelike case
this concept is not gauge invariant.\\
We now want to divide out by the gauge transformations generated by 
$U(*)\;\in\;SU(1,1)$ to obtain the physically relevant range of the 2 real 
numbers $a_i$. Given a pair $a_i,t^I$ one can show that by a SU(1,1) gauge 
transformation one can get another pair $b_i,s^I$ such that in case that
$t^I$ is\\
a) timelike : $b_i=a_i\mbox{ and }t^I$ is any normalized future directed
timelike vector,\\
b) spacelike : $b_i=\pm a_i\mbox{ and }t^I$ is any normalized spacelike 
vector and\\
c) null : $b_i=s a_i\mbox{ and }t^I$ is any future directed null vector.\\
So we see that $t^I$ is pure gauge up to its causal characterization and
its normalization and that in the timelike, spacelike
and null regime repectively, the space coordinatized by the $a_i$ is 
topologically  
$S^1\times S^1,\;(R^1\times R^1)/Z_2
\mbox{ and } S^1$ respectively (to see this in the timelike case, observe
that the holonomy is given then up to a sign in terms of sines and cosines).\\
Another way to see this is by explicit calculation of the exponential part of 
the homomorphism 
\beq
\exp(a\tau_I t^I)=\left \{ \begin{array}{r@{\quad:\quad}l}
\mbox{cosh}(a/2)1+t^I\tau_I\mbox{sinh}(a/2) & \mbox{spacelike sector} \\
\mbox{cos}(a/2)1+t^I\tau_I\mbox{sin}(a/2) & \mbox{timelike sector} \\
1+t^I\tau_I a & \mbox{null sector}
\end{array} \right \} \eeq
(where we may think of a as $a=a_i\oint_\alpha dx^i$ in the above mentioned
gauge). 
Taking the trace we
infer that the gauge invariant information captured by the exponential part
of the holonomy is given by, for the spacelike
sector in $a_i\;\in\; R^2/Z_2$, for the timelike sector in $a_i\;\in\; T^2$
and
for the null sector only the angle between $a_1,a_2$ is gauge invariant since
$t^I$ gets not only rotated but also scaled by a positive scale factor
so that the reduced configuration space has then the topology of the
circle $S^1$. Note that $T^0$ does not capture this piece of information
in the null regime (\cite{5}). We are not interested in the null regime in the 
sequel.\\
Now, let us invoke the piece of information that is captured by the sign of the
homomorphism in the non timelike case. Obviously, 4 distinct 
assignments of signs to the 2 generators are possible, so we get 4 distinct 
copies 
of either $R^2/Z_2\mbox{ or }S^1$. These 4 copies are disconnected since there 
is no continous way to get, say, a boost of type (+,+) from a boost of type
(-,+).\\
Finally, we should observe that the timelike sector is connected to the
two other causal sectors at the zero connection. Hence this special point
does not correspond to either of these sectors since the causal nature of the
zero connection is degenerate. So one should discard this point from either
of the three causal sectors to get finally the topology of the reduced 
configuration space as :\\
timelike : $T^2-{0}$, spacelike : $(R^2/Z_2-{0})\times Z_4\approx(R^2-{0})
\times Z_4$, null : $(S^1-{0})\times Z_4$ and\\
 zero : ${0}$.
Thus, the reduced phase space is just the cotangent bundle over the disjoint
union of these sectors of the reduced configuration space.\\
An interesting question is now which of these sectors corresponds to
geometrodynamics, that is, such that the intrinsic 2-metric is nondegenerate
and has euclidean signature.\\
Here it should be stressed that the Witten formulation of 2+1 gravity differs
tremendously from the geometrodynamical formulation :
For instance, Witten shows (\cite{1}) that we can write the 
action (2.2) as the action
of a ISO(1,2) or ISU(1,1) Chern-Simons field theory whose connection
$\omega_I=A_a^I\tau_I+e_a^I T_I$ ($T_I$ are the generators of the translation
subgroup of the Poincar\'e subgroup) is flat on shell. That means locally we
can always gauge $\omega$ and in particular the triad to zero, corresponding to
a vanishing (!) metric. In other words, the Witten constraints generate more
general gauge transformations than only spacetime diffeomorphisms and 
issues like degeneracy and signature of the induced hypersurface become
gauge dependent.\\
At this point the paper by Mess (\cite{18}) should be mentioned. Mess shows 
that if one starts from the geometrodynamical formulation for the torus,
i.e. a nondegenerate metric on a spacelike hypersurface, then the 
holonomies are either boosts or the identity map, which means that the 
associated connection is either spacelike or zero.
The interested reader is 
referred to that paper and also to the papers by Louko and Marolf (\cite{7}).

\subsection{Quantization}

As outlined in reference \cite{7}, the quantization for the timelike sector
is much easier than for the spacelike and null sector. In fact, it turns out 
that the kernel of the naively defined loop transform (compare Br\"ugmann's 
lectures) for the spacelike sector is dense in the
Hilbert space that is defined by the loop algebra. Although the
authors of that paper suggest a way how to define a loop transform for the
spacelike sector such that it becomes an isomorphism between
the loop representation and the connection representation\footnote{essentially,
these authors show that there is a dense subspace of the connection
representation w.r.t. which the naively defined loop transform (compare also
(2.21)) is faithful; they then Cauchy complete this image w.r.t. the inner
product that coincides with that for the connection representation in the
pre-image; the two representations are thus unitarily equivalent by 
construction, for still another procedure refer to \cite{16}}, the techniques
involved would veil the main ideas of the formalism and so we stick to the
timelike sector. This, again, does not simulate the 3+1 case because there the
$T^0$'s are unbounded.\\

Let us start with the construction of the loop
representation. The idea of the loop representation is to use
a {\em non-canonical} subalgebra of the Poisson-algebra as the basic set of
classical variables that are to be quantized by demanding that the
commutation relations among these variables (which in case of the loop
representation are just the $T^0,T^1$'s) mirror those of the classical
analogues. By doing that, one gets rid of the connections and the electric
fields. They should, however, be reconstructed by means of the so-called
loop transform (refer, for example, to \cite{19} for more details). Let us now 
make these ideas concrete.\\
First, we need to compute the classical Poisson algebra among the $T^0,T^1$
induced by that for $E^a_I,A_a^I$. We obtain
\begin{eqnarray}
\{T^0_\alpha,T^0_\beta\} & = & 0 \nonumber\\
\{T^0_\alpha,T^1_\beta\} & = & \sum_i \Delta_i(\alpha,\beta)
[T^0_{\alpha\circ_i\beta}-T^0_{\alpha\circ_i\beta^{-1}}] \nonumber\\
\{T^1_\alpha,T^1_\beta\} & = & \sum_i \Delta_i(\alpha,\beta)
[T^1_{\alpha\circ_i\beta}-T^1_{\alpha\circ_i\beta^{-1}}]
\end{eqnarray}
where i labels points of intersection of the two loops involved, $\circ_i$
denotes composition at the intersection point and we have defined the
quantity $\Delta_i$ which takes values in $\{\pm 1\}$ by
\beq \sum_i \Delta_i(\alpha,\beta):=\int_{\alpha\times\beta} dx\wedge dy
\delta^{(2)}(x,y)\;. \eeq
Note also that the T variables are classically real.\\
We now quantize this algebra by demanding that the commutators among
the T's produce $i\hbar$ times the right hand side of the Poisson brackets
and by imposing the following *-relations with respect to an abstract
involution * : $(\hat{T}^0)^*=\hat{T}^0,\,(\hat{T}^1)^*=\hat{T}^1$.
We choose the loop representation, that is, we represent the T operators
on a complex vector space of functions f of loops. Secretely, one should
think of such a function as
arising from a function $\tilde{f}$ of connections via a heuristic loop
transform
\beq f(\{\beta\})=\int_{{\cal A}/{\cal G}} d\mu[A]\prod_{\beta\in\{\beta\}}
T^0_\beta[A]\tilde{f}(A)
\eeq
where $\{\beta\}$ is a set of single loops $\beta$
and the integration domain is the moduli space ${\cal A}/{\cal G}$ of (flat)
connections modulo gauge transformations.\\
The analysis is now simplified by the fact that for any subgroup of SL(2,C)
that is generated by some real form of sl(2,C) there holds the SL(2,C) 
Mandelstam identity
\beq T^0_{\alpha\circ\beta}+T^0_{\alpha\circ\beta^{-1}}
=T^0_{\alpha}\hat{T}^0_{\beta} \eeq
which implies that any analytic function of traces of holonomies can be
written as a linear combination of traces of holonomies for single loops.
We therefore need only to define the action
of our loop operators on functions of single loops :
\beq f(\alpha)=\int_{{\cal A}/{\cal G}} d\mu[A] T^0_\alpha[A]\tilde{f}(A)\;.
\eeq
It turns out that one can implement the algebra (2.17) then as follows
\begin{eqnarray}
(\hat{T}^0_\alpha f)(\beta):=f(\alpha\circ\beta)+
f(\alpha\circ\beta^{-1}) \nonumber\\
(\hat{T}^1_\alpha f)(\beta):=i\hbar\sum_i \Delta_i(\alpha,\beta)
[f(\alpha\circ_i\beta)-f(\alpha\circ_i\beta^{-1})]
\end{eqnarray}
where the composition $\circ$ is at the basepoint.\\
In order to make sure that the T observables 'come from a connection', we
have to impose further the following identities on our representation space
\begin{eqnarray}
i) & &  \hat{T}^1_\alpha=\hat{T}^1_{\alpha^{-1}},\;
\hat{T}^1_{\alpha\circ\beta}=\hat{T}^1_{\beta\circ\alpha} \nonumber\\
ii) & & \hat{T}^0_{\alpha\circ\beta}+\hat{T}^0_{\alpha\circ\beta^{-1}}
=\hat{T}^0_{\alpha}\hat{T}^0_{\beta} \nonumber\\
iii) & & \hat{T}^0_*=2,\;\hat{T}^1_*=0 \;.
\end{eqnarray}
It is easy to check that the analogue of i) for $T^0$ follows from the
SL(2,C) Mandelstam identity ii).\\
Thus we restrict the representation space further such that the relations
i)-iii) hold. One can check that it is enough to demand that ($a_i$ are real
quantities)
\beq \sum_i a_i f(\alpha_i\circ\beta)=0\mbox{ whenever }
\sum_i a_i T^0_{\alpha_i}=0. \eeq
This, again, follows trivially from the existence of a loop transform 
(2.19).\\
Next, we restrict the loop algebra and its representation space to depend
only on homotopy classes. For the torus this amounts to the fact that
T-operators and the functions of our complex vector space depend
only on the winding numbers $n_1,n_2\in Z$ of the 2 generators of the
abelian fundamental group of $\Sigma$. If we denote by $[\alpha_i],\;i=1,2$
the generators of $\pi_1(\Sigma)$ then the action (2.22) becomes
\begin{eqnarray}
(\hat{T}^0_{\alpha_1}f)(n_1,n_2)=f(n_1+1,n_2)+f(n_1-1,n_2) \nonumber\\
(\hat{T}^1_{\alpha_1}f)(n_1,n_2)=i\hbar n_2[f(n_1+1,n_2)-f(n_1-1,n_2)]
\end{eqnarray}
where we have made use of the relation
\beq f(n_1,n_2)=f(-n_1,-n_2),\; (n_1,n_2)\sim(n_1',n_2')\Leftrightarrow
(n_1',n_2')=(-n_1,-n_2) \eeq
which follows from (2.23). The action of the T operators for the other 
homotopy class is analogous. The reason for the factor of $n_2$ in the second
line of (2.25) is that any representative of $[\alpha_1]$ will intersect any
representative of $[\alpha_1]^{n_1}[\alpha_2]^{n_2}$ precisely $|n_2|$ times
with the orientation of this intersection captured by the sign of $n_2$.\\
The final task is now to find an inner product such that our basic operators
become self-adjoint. The obvious choice is
\beq <f,g>:=\sum_{Z^2/\sim} \bar{f}(n_1,n_2)g(n_1,n_2) \eeq
and obviously accomplishes our aim.\\

\subsection{Loop transform}

In order to explicitely construct the loop transform of the present model,
we need also the connection representation.\\
Physical states depend only on the moduli space of flat connections modulo
gauge transformations, labelled by the two parameters $a_1,a_2\in
[-2\pi,2\pi]$. We realize our basic operators in the connection
representation as follows
\begin{eqnarray}
(\hat{T}^0_{\alpha_i}\tilde{f})(a_1,a_2) &:=& 2\cos(a_i/2)\tilde{f}(a_1,a_2)
\nonumber\\
(\hat{T}^1_{\alpha_i}\tilde{f})(a_1,a_2) &:=& -4i\hbar\sin(a_i/2)
\frac{\partial\tilde{f}}{\partial a_j}(a_1,a_2) \;,i\not=j
\end{eqnarray}
and it is straightforward to check that the commutation relations (2.17)
are satisfied.\\
An inner product that makes these operators
self-adjoint can be constructed as follows : we make the ansatz
\beq (\tilde{f},\tilde{g}):=\int_{[-2\pi,2\pi]^2}da_1\wedge da_2
\rho(a_1,a_2)\bar{\tilde{f}}(a_1,a_2)\tilde{g}(a_1,a_2) \eeq
and try to find a weight factor $\rho$ such that the T operators are
self-adjoint. We find up to a multiplicative positive constant $\rho=1$.\\
Alternatively we may apply the following procedure which is appropriate
whenever one is confronted with a non-canonical algebra of basic operators
(the interested reader is referred to \cite{9}) :\\
Choose some Riemannian background metric on the torus and choose physical
states as densities of weight $1/2$. Define the vector fields
\beq v_i:=4\sin(a_i/2)\frac{\partial}{\partial a_j} \eeq
and define the action of the T observables as in (2.28) on half-densities
except that $\hat{T}^1_{\alpha_i}\tilde{f}:=-i\hbar{\cal L}_{v_i}\tilde{f}$
where $\cal L$ denotes the Lie derivative.\\
It is then easy to check that the scalar product (2.29) with $\rho=1$
is well-defined (that is, frame independent) and that all operators are
self-adjoint with respect to it.\\
We are now finally in the position to look at the loop transform. Taking
the Lebesgue measure on the torus as the measure $\mu$ in (2.21) we obtain
\beq f(n_1,n_2)=\int_{[-2\pi,2\pi]^2}d^2a\cos([n_1 a_1+n_2 a_2]/2)
\tilde{f}(a_1,a_2) \eeq
and one can explicitely check that any of the elementary operators $\hat{O}$
in (2.25) for the loop representation arises from the operators
$\hat{\tilde{O}}$ in (2.28) for the connection representation via the loop
transform, that is
\beq (\hat{O}f)(\alpha)=\int_{{\cal A}/{\cal G}} d\mu(A) T^0_\alpha(A)
(\hat{\tilde{O}}\tilde{f})(A)\;. \eeq
Note that the space of functions which are either odd or
even under reflection of the $a_i$ is left invariant under the action of
the T operators, so the representation space splits into 2 irreducible
representations. It follows that the loop transformation (2.31) is a
representation isomorphism for the even sector and has the odd sector
as its kernel. Since for the even sector (2.31) is just the Fourier transform,
it follows that the Hilbert spaces that we have constructed are unitarily
equivalent, that is $<f,g>=(\tilde{f},\tilde{g})$.

\subsection{Discussion}

The model of 2+1 pure gravity is a dimensional reduction of 3+1 gravity.
It can be cast into a form such that the algebraic structure of the
constraints is very similar to that of the full theory. The gauge group,
SU(1,1), is not compact and therefore simulates the fact that the
traced holonomy for the complex Ashtekar connection in 3+1 gravity is an 
unbounded function (in the topology of complex numbers).
Both, the connection representation and the loop representation,
together with the loop transform can be constructed in closed form. This
gives one some confidence that a loop representation of the 3+1 theory can
be constructed as well.\\
These were the positive remarks. However, sharp criticism is in order :\\
The 2+1 theory is manifestly real whereas the 3+1 case is genuinely
complex. Since it is fair to say that some of the major difficulties
of the physical theory rest in the non-trivial reality structure, one
should not expect that the constructions that were used in the model are
available in the 3+1 case. This has nothing to do with the fact that we only 
quantized the sector in which
the $T^0$'s are bounded because, as mentioned, one {\em can} take care of
the spacelike sector. Rather what we point out here is that a genuinely
complex connection gives rise to additional difficulties, for example the
$T^0,\; T^1$ Poisson algebra is not even a *-algebra in this case. \\
Next, in the above analysis we restricted ourselves to the Witten sector.
Thus, we did not use the form of the constraints that actually equals that
of 3+1 gravity. The connection of 3+1 gravity is not flat in general, the
theory is not topological and, most seriously, maybe the major difficulty
of the full theory is
that the scalar constraint is {\em quadratic} in the momenta. However, we
used only constraints that are linear in or independent of the momenta.
Accordingly, 2+1 gravity neither accounts for this issue.\\
Last but not least, the fact that the reduced phase space is finite
dimensional is a severe deviation from the 3+1 case because all the
divergencies of quantum field theory occur due to the infinite number of
degrees of freedom. We were thus unable to deal with this technical problem.

\section{Spherically symmetric gravity}

\subsection{Canonical formulation}

The model of spherically symmetric gravity arises from a Killing reduction
of 3+1 gravity. We will follow largely reference \cite{10} in the sequel.\\
We first summarize the Killing reduction first given in \cite{11}
\footnote{In order to make contact with ref. \cite{11}, one has to exchange
the labels I=2 and I=3 and to replace $A_3+\sqrt{2}$ there by $A_3$ in order
to get $A_3$ here}.\\
One imposes spherical symmetry on the geometrodynamical phase space by
requiring that the induced metric on the Cauchy hypersurface $\Sigma$ 
(foliated by SO(3) orbits) and
its extrinsic curvature be Lie-annihilated by three Killing vector fields
$K_i$ which form an SO(3) subalgebra of the Lie algebra of all vector fields
on $\Sigma$. In case of the Ashtekar phase space, one has to look for
analogous conditions for the triad and the Ashtekar connection. Let us
start with the triad. We have
\[ {\cal L}_{K_l} q_{ab}=2\delta_{ij}e^i_{(a}{\cal L}_{K_l}e^j_{b)}=0 \]
and the most general solution of this equation is given by
\beq {\cal L}_{K_l}e^i_a=\epsilon^i\;_{lk}e^k_a\;, \eeq
that is, a rotation of the triad in the tangent space in either of the three
Killing directions is compensated
by a corresponding rotation in the internal space.\\
The idea is now as follows : obtain the general solution of (3.1). This is
then the general form of a 1-form that transforms according to
the defining representation of SO(3). Since the defining and the adjoint 
representation of SO(3) are isomorphic, the solution of (3.1) is also the 
general form of an so(3)-valued 1-form. Finally, since the three infinitesimal
internal rotations parametrized by $\Lambda^{(l}_i=\delta^l_i$
involved in (3.1) are global ones, the inhomogenous term in the  
transformation law of the (pull-back to the base manifold of a) connection
$\omega_a^i\;\to\;-\partial_a \Lambda^i+\epsilon_{ijk} \Lambda^j \omega_a^k$
drops out so that the solution to (3.1) also gives us the general form of
a SO(3)-connection.
To obtain then the general form of
a SO(3) valued vector or vector density, one simply contracts the solution
to (3.1) with $q^{ab}$ times an appropriate power of $\sqrt{\det(q)}$ since
the three-metric is Lie-derived by the Killing-fields. The
form of $q_{ab}$ is already known from the solution to (3.1).\\
We will shortly sketch the calculations. Fix a local frame $(r,\theta,\phi)$
on $\Sigma$ where r is a local radial coordinate and $\theta,\phi$ are
global angular coordinates. We can then write the rotational Killing fields
as follows (compare any textbook in quantum mechanics) :
\begin{eqnarray*}
-K_1 & = & -\sin(\phi)\partial_\theta-\cot(\theta)\cos(\phi)\partial_\phi \\
-K_2 & = &  \cos(\phi)\partial_\theta-\cot(\theta)\sin(\phi)\partial_\phi \\
-K_3 & = & \partial_\phi
\end{eqnarray*}
and plugging these expressions into (3.1) we obtain a system of partial
differential equations of first order. One solves it for the partial
derivatives of $e_a^i\mbox{ wrt }\theta,\phi$ and after some algebraic
manipulations one sees that one can decouple the equations. The final,
unique solution is then
\[ (e_r^i,e_\theta^i,e_\phi^i)=(f n_r^i,g n_\theta^i+h n_\phi^i,
g n_\phi^i-h n_\theta^i) \]
where $n_a^i$ is the standard orthonormal base in internal space
(that is\newline $n_r^i=(\sin(\theta)\cos(\phi),\sin(\theta)\sin(\phi),
\cos(\theta)),\;n_\theta^i=\partial_\theta n_r^i,\;\sin(\theta)n_\phi^i
=\partial_\phi n_r^i$) and f, g and h are arbitrary real functions of the
radial (r) and time (t) variable only. From this we conclude, as expected,
the general form of the 3-metric
\beq q_{ab}=q_{rr}(r,t) r_{,a}r_{,b}+q_{\theta\theta}(r,t)h_{ab} \eeq
where $h_{ab}$ is the standard line element on the 2-sphere.\\
Inverting (3.2), computing $E^a_i=\sqrt{\det(q)}q^{ab}e_b^i$ we see that
we can parametrize the general form of the conjugate variables of the
Ashtekar phase space as follows (the numerical factors are chosen for
convenience)
\begin{eqnarray}
& & (E^r_i,E^\theta_i,E^\phi_i)=
(E^1 n_r^i\sin(\theta),\frac{\sin(\theta)}{\sqrt{2}}(E^2
n_\theta^i+E^3n_\phi^i),\frac{1}{\sqrt{2}}(E^2 n_\phi^i-E^3 n_\theta^i)) \\
& & (A_r^i,A_\theta^i,A_\phi^i)=(A_1n_r^i,\frac{1}{\sqrt{2}}(A_2
n_\theta^i+(A_3-\sqrt{2}) n_\phi^i), \\  & & ~~~~~~~~~~~~
\frac{\sin(\theta)}{\sqrt{2}}(A_2 n_\phi^i-(A_3-\sqrt{2})
n_\theta^i)) \; .\nonumber
\end{eqnarray}
The arbitrary complex functions
$E^I=E^I(t,r),\;A_I=A_I(t,r); I=1,2,3$ depend on t and r only. One now
simply inserts this into Ashtekar's action of full gravity (compare
Giulini's lectures), integrates out the angles (in particular the factor
$\sin\theta$ contained in $\tiN$ drops out) and finally ends up with the
following action
\beq S=\frac{4\pi}{\kappa}\int_R dt(\int_\Sigma dr[-i\dot{A}_I E^I-[i\Lambda
{\cal G}-iN^r V +\tiN C]]+b) \;. \eeq
The explanation of the various quantities involved in (3.5) is as follows :\\
All quantities only depend on r,t (expanding all quantities in terms of
spherical functions, only the angle-independent terms survive after
integrating over $S^2$). The Lagrange multipliers and constraint functions
are related to those of the full theory by $\Lambda=\Lambda^i (n_r)^i,\;
\sin(\theta){\cal G}={\cal G}_i (n_x)^i,\;
N^r=N^a r_{,a},\;\sin(\theta)V=V_a r_{,a},\;\tiN=\sin(\theta)\tiN_{full},\;
2\sin^2(\theta)C=C_{full};\; \kappa/8\pi$ is Newtons constant.\\
The constraint functionals take the following form
\begin{eqnarray}
& & {\cal G}=(E^1)'+A_2 E^3-A_3 E^2\mbox{ : Gauss constraint,}\\
& & V=B^2 E^3-B^3 E^2\mbox{ : Vector constraint,}\\
& & C=\frac{1}{2}(E^2(2 B^2 E^1+B^1 E^2)+E^3(2 B^3 E^1+B^1E^3))\mbox{ :
Scalar constraint.}
\end{eqnarray}
Here we have used the reduction to spherical symmetry of the magnetic
fields $B^a_i=1/2\epsilon^{abc}F_{ab}^i$ (where F denotes the field strength
of the Ashtekar connection)
\beq (B^r_i,B^\theta_i,B^\phi_i)=
(B^1 n_r^i\sin(\theta),\frac{\sin(\theta)}{\sqrt{2}}(B^2
n_\theta^i+B^3n_\phi^i),\frac{1}{\sqrt{2}}(B^2 n_\phi^i-B^3 n_\theta^i))
\eeq
and one can check that
\beq (B^1,B^2,B^3)=(\frac{1}{2}((A_2)^2+(A_3)^2),(A_3)'+A_1 A_2,
-(A_2)'+A_1 A_3) \eeq
where a prime denotes differentiation with respect to r.\\
Since we are interested in asymptotically flat topologies in contrast to
the first model, there is also a boundary term b involved in the action
which is to make the action functionally differentiable. It reads
\beq b=\int_{\partial\Sigma} i N^r(A_2 E^2+A_3 E^3)+\tiN (A_2
E^3-(A_3-\sqrt{2})E^2)E^1 \eeq
and can be recognized as the sum of ADM momentum and energy. It is also
obvious from the action (3.5) that $A_I, E^I$ form a canonical pair. With
this boundary term we are able to derive the following equations of motion
\begin{eqnarray}
\frac{d}{dt}A_1 & = & i[(-i\Lambda'+\tiN(B^2 E^2+B^3 E^3)],\\
\frac{d}{dt}A_2 & = & i[(-i\Lambda A_3+i N^r B^3+\tiN(B^2 E^1+B^1 E^2)],
\nonumber\\
\frac{d}{dt}A_3 & = & i[(+i\Lambda A_2-i N^r B^2+\tiN(B^3 E^1+B^1 E^3)],
\nonumber\\
\frac{d}{dt}E^1 & = & -i[-i N^r(A_2 E^3-A_3 E^2)+\tiN (A_2 E^2+A_3 E^3)E^1],
\nonumber \\
\frac{d}{dt}E^2 & = & -i[+i(\Lambda-N^r A_1)E^3+i(N^r E^2)'+\tiN(A_1 E^1E^2
+\frac{1}{2}E A_2) \nonumber \\
&  & +(\tiN E^1 E^3)'], \nonumber\\
\frac{d}{dt}E^3 & = & -i[-i(\Lambda-N^r A_1)E^2+i(N^r E^3)'+\tiN(A_1
E^1E^3+\frac{1}{2}E A_3) \nonumber \\&  & -(\tiN E^1 E^2)']\; .\nonumber
\end{eqnarray}
We can also display the classical canonical constraint algebra (we
abbreviate $f\circ g:=\int_\Sigma dr f(r) g(r)\;, \xi:=V-A_1{\cal G}$) :
\begin{eqnarray}
\{\Lambda_1\circ{\cal G},\Lambda_2\circ{\cal G}\}&= & 0,  \\
\{N\circ\xi,\Lambda\circ{\cal G}\} & = &-i
N\Lambda'\circ\cal G, \nonumber\\
\{\tiN \circ H,\Lambda\circ{\cal G}\} & = &0, \nonumber\\
\{M\circ\xi,N\circ\xi\} & =& i(M N'-M'N)\circ\xi, \nonumber \\
\{M\circ\xi,\tiN\circ H\}
& = & i(M\tiN'-M'\tiN)\circ H, \nonumber \\
\{\tiM\circ H,\tiN\circ H\} & = &
i(\tiM\tiN'-\tiM'\tiN)\circ(E^1)^2 H_x  \nonumber
\end{eqnarray}
and it is obvious that the model is still first class (recall Wipf's
lectures).\\
The set of equations (3.12) and (3.13) allow for the following
interpretation :\\
The diffeomorphisms have been  frozen to the r-direction, the
internal rotations to the $n_r$-direction. $A_1$ plays the role of an $O(2)$
gauge potential, $E^1$ is $O(2)$-invariant while the vectors $(E^2,E^3),
(A_2,A_3)$ transform
according to the defining representation of $O(2)$. $A_1,E^2,E^3$ are
densities of weight one in one dimension, while $E^1,A_2,A_3$ are scalars.\\
For the discussion of the reality  conditions
the reduction to spherical symmetry of the spin connection is
needed:
\begin{eqnarray}
(\Gamma_r^i,\Gamma_\theta^i,\Gamma_\phi^i) & = &(\Gamma_1
n_r^i,\sqrt{\frac{1}{2}}(\Gamma_2 n_\theta^i+(\Gamma_3-\sqrt{2})n_\phi^i),\\
& & ~~~~~~~~~~~~
\sqrt{\frac{1}{2}}(\Gamma_2 n_\phi^i-(\Gamma_3-\sqrt{2})n_\theta^i)
\sin(\theta)),\nonumber \\
(\Gamma_1,\Gamma_2,\Gamma_3) & = & (-\beta',-(E^1)'\frac{E^3}{E},(E^1)'
\frac{E^2}{E}) \;
,~~~\beta'= \frac{E^2 E^{3 \prime} -E^3 E^{2\prime}}{E} \; \end{eqnarray}
and we have introduced the following quantities :
\beq
\alpha:=\arctan(\frac{A_3}{A_2}),~~~\beta:=\arctan(\frac{E^3}{E^2}),~~~
A:=(A_2)^2+(A_3)^2,~~~E:=(E^2)^2+(E^3)^2. \eeq
The metric is given, in these variables, by
\beq q_{ab}=\frac{E}{2E^1} r_{,a}r_{,b}+E^1 h_{ab} \;.\eeq
Then the reality conditions become simply
\beq E^I=\mbox{real and }A_I-\Gamma_I=\mbox{imaginary.} \eeq
To complete the Hamiltonian formulation we have to agree on the topology
of the 1-dimensional hypersurface $\Sigma$ as well as on the fall-off
properties of the fields at spatial infinity.\\
First note that the 3-dimensional hypersurface is related to our
1-dimensional one by $\Sigma^{(3)}=\Sigma\times S^2$. Now we can choose
$\Sigma$ either to be closed or open. That is, we choose its topology to
be either $S^1$, the circle, or $R^1$, the real line. In the first case
we are dealing with a compactified wormwhole, in the second with a black
whole with two asymptotic regions. One can, at the price of inventing
additional assumptions, generalize to more than 2 asymptotic ends (\cite{10})
but we refrain from doing that here for the sake of brevity.\\
In the asymptotically flat case we also have to deal with boundary conditions
on the fields.\\
First, we choose the hypersurface label $r\in[-\infty,\infty]$
to become asymptotically $r^2=(x^1)^2+(x^2)^2+(x^3)^2$ with respect to an
asymptotical cartesian frame $\{x^a\}$ which means that it is appropriate
to describe fall-off properties in powers of r.\\
Next, following reference \cite{12}, we adopt the requirements of that
paper to our situation. The guiding principle for choosing fall-off
properties are :\\
1) finiteness of the symplectic structure, \\
2) finiteness and functional differentiability of the constraint functionals.\\
Requirement 2) further depends on the set of asymptotic symmetries that one
is willing to allow.\\
In \cite{12} (which is based on the old (ADM) variables) these requirements
1) and 2) including asymtotical Poincare transformations can be satisfied as
follows :\\
\begin{eqnarray}
q_{ab}&\rightarrow& \delta_{ab}+\frac{f_{ab}(x^c/r,t)}{r}+O(1/r^2)
\nonumber\\
p^{ab}&\rightarrow& \frac{k^{ab}(x^c/r,t)}{r^2}+O(1/r^3)
\end{eqnarray}
as $r\rightarrow\infty$. Furthermore, it must be
required that the functions $f_{ab}$ and $k^{ab}$ respectively are even and
odd respectively under reflections of the asymptotically flat frame.\\
It is clear that for spherical symmetry we are not able to impose the above
parity conditions because the reduction to spherical symmetry excludes
all modes of the fields (regarded as expanded into spherical harmonics)
which have angular momentum different from zero. Hence we have to modify
the strategy slightly.\\
Comparing the spherically symmetric metric (3.17)
with the Euclidean metric in spherical coordinates
\beq \delta_{ab}=r_{,a}r_{,b}+r^2 h_{ab} \; , \eeq
we conclude the following fall-off properties :
\begin{eqnarray}
 & & (E^1,E^2,E^3)\rightarrow (r^2[1+\frac{f^1(t)}{r}+O(1/r^2)], \nonumber\\
 & & \sqrt{2}r[\bar{E}^2+
\frac{f^2(t)}{r}+O(1/r^2)],\sqrt{2}r[\bar{E}^3+\frac{f^2(t)}{r}+O(1/r^2)])
\end{eqnarray}
whereby $(\bar{E}^2)^2+(\bar{E}^3)^2=1$. Inserting this into the formula
$A_a^i=\Gamma_a^i+iK_a^i$ (recall Giulini's lecture), using (3.15), (3.19)
and (3.3) one concludes that
\begin{eqnarray}
(A_1,A_2,A_3-\sqrt{2}) & \rightarrow &(\frac{a_1(t)}{r^2}+O(1/r^3),
\frac{a_2(t)}{r}+O(1/x^2),\frac{a_3(t)}{r}+O(1/r^2)) \; .
\end{eqnarray}
Since, as we noted before, there is no parity freedom left, the requirements
1) and 2) discussed above will not be satisfied yet. Let us explore what
further restrictions are there to be imposed.\\
The symplectic structure on the large phase space can be read off from the
action (we can drop the prefactor of the action for the case of pure
gravity) :
\begin{eqnarray}
& &  \Omega=\int_\Sigma dr [-i dE^I\wedge dA_I]
\nonumber\\
& &=\int_\Sigma dr (\frac{-i}{r}[da_1\wedge df^1+\sqrt{2}(da_2\wedge
df^2+da_3\wedge df^3)]+O(1/r^2)) \; .
\end{eqnarray}
Hence we can satisfy requirement 1) by restricting the variations to be such
that
\beq da_1\wedge df^1+\sqrt{2}(da_2\wedge df^2+da_3\wedge df^3)=0 .\eeq
As for requirement 2) we first have to agree on the set of allowed symmetries
at infinity. We want to incorporate only asymptotic translations. Why do we
not consider asymptotic boosts of the 2-dimensional flat structure (rotations
do not exist in 1 dimension anyway) ? In the literature, one looks at
Schwarzschild-solutions in arbitrarily boosted frames (see ref. \cite{12}, for
example). However, these boosts are really boosts with respect to the
4-dimensional spacetime which violate spherical symmetry of the initial data.
The 'boosts' that we were able to discuss here must be meant with respect to
the effective 2-dimensional spacetime coordinatized by the variables r and t
in order not to violate spherical symmetry, they are thus not physical
anyway. But since we do not have this parity freedom at our disposal our
'boost' generator diverges. So we would have to impose much more
restrictive fall-off conditions than above which, in particular, would
exclude Schwarzschild configurations and for that reason we refrain
from doing so.\\
The same is actually also true for asymptotic spatial translations : only 
radial translations preserve the spherical symmetry of the fields, that is,
translations of the form $x^a\rightarrow x^a+c x^a/r$ where c is a constant
but these are then position-dependent (on the sphere) and do not correspond
to the translation subgroup of the Poincar\'e group. However, we will
keep them for completeness sake.\\
Obviously, we have then {\em for symmetry transformations} the following
fall-off behaviour of the Lagrange multipliers :
\beq
(\Lambda,N^x,\tiN)\rightarrow(\frac{\mbox{const.}}{r^2}+O(1/r^3),
\mbox{const.}
+O(1/r),\frac{\mbox{const.}}{r^2}+O(1/r^3))
\eeq
while {\em for gauge transformations} we require, for simplicity, that the
Lagrange multipliers are of compact support.\\
We now compute the leading order behaviour of the integrands of the
constraint functionals :
\beq ^E{\cal G}\rightarrow 2r(1-\bar{E}^2)+f^1+\sqrt{2}(a_2\bar{E}^3-
\sqrt{2}f^2-\bar{E}^3 a_3)+O(1/x) \eeq
which becomes a finite and differentiable functional when imposing
$\bar{E}^2=1$ i.e. $\bar{E}^3=0$.\\
Note that weakly (i.e. on the constraint surface) we have from the Gauss
constraint at infinity
\beq f^1-2f^2-\sqrt{2}a_3=0 \; . \eeq
It is convenient first to compute the asymptotic form of the
magnetic fields
\begin{eqnarray}
B^1 &\rightarrow& -\frac{\sqrt{2}a_3}{r}+O(1/r^2), \nonumber \\
B^2 &\rightarrow& -\frac{a_3}{r^2}+O(1/r^3), \nonumber \\
B^3 &\rightarrow& \frac{a_2+\sqrt{2}a_1}{r^2}+O(1/r^3)
\end{eqnarray}
to conclude for the vector constraint
\beq V\rightarrow\sqrt{2}\frac{a_2+\sqrt{2}a_1}{r}+O(1/r^2) \; .\eeq
Hence we have to impose
\beq a_2+\sqrt{2}a_1=0 \eeq
in order to make this functional
finite and differentiablility can be achieved by adding the ADM-momentum.
Finally, it is easy to see that with this restriction the scalar
constraint functional is already finite and functionally differentiable
when adding the ADM-energy.\\
Now it is possible to make the restriction that comes from requirement 1)
more concrete. We have
\begin{eqnarray}
0 & = & da_1\wedge df^1+\sqrt{2}(da_2\wedge df^2+da_3\wedge df^3)\nonumber\\
  & = & -\frac{1}{\sqrt{2}}da_2\wedge d(f^1-2f^2)+\sqrt{2}da_3\wedge df^3)
      \nonumber\\
  & = & -\frac{1}{\sqrt{2}}da_2\wedge d(f^1-2f^2-\sqrt{2}a_3)
        +da_3\wedge d(\sqrt{2}f^3+a_2) \; .
\end{eqnarray}
Note that the bracket of the 1st wedge product in the last line of (3.31)
vanishes weakly according to (3.27). Hence it is consistent with the
constraint equations to impose
\beq \sqrt{2}f^3+a_2=0\; . \eeq
This completes the analysis of the boundary conditions.\\
It is clear that the present model bears a strong resemblance to the full
theory : the algebraic structure of the constraints is very similar, the
reality conditions are non-trivial, the constraints are quadratic in the
momenta.

\subsection{Symplectic reduction}

Let us first recall some basic facts from the theory of symplectic reduction
(for an extensive treatment, see ref. \cite{13} and \cite{14}).\\
We showed in the previous section that the present model is a field theory
with first class constraints. Let $\Gamma,\bar{\Gamma}\;\mbox{and}\;
\hat{\Gamma}$ denote
the full phase space, its constraint surface (where the constraints are
identically satisfied) and its reduced phase space (i.e. the constraint
surface, but points in it are identified provided they are gauge
related). The (local) existence of the latter follows from general theorems
that are valid for first class systems. Let
\beq \iota\; :\; \bar{\Gamma}\rightarrow\Gamma\;\mbox{and}\;\pi\; :\;
\bar{\Gamma}\rightarrow\hat{\Gamma} \eeq
denote the (local) imbedding and projection into the large phase space and
onto the reduced phase space respectively. Call the symplectic structures
on the 2 phase spaces $\Omega\;\mbox{and}\;\hat{\Omega}$ respectively. Then
the presymplectic structure on $\bar{\Gamma}$ is defined by the pull-backs
\beq \pi^*\hat{\Omega}:=\bar{\Omega}:=\iota^*\Omega \;. \eeq
(More precisely, in practise one computes the constraint surface and thus
obtains the imbedding. One then defines the presymplectic structure by the
pull-back under the imbedding. After that one computes the gauge orbits
and obtains the projection. The reduced symplectic structure is then
defined by the pull-back under the projection).\\
On the other hand, if $\Theta\;\mbox{and}\;\hat{\Theta}$ are the symplectic
potentials for the symplectic structures, we obtain
\beq d\wedge(\iota^*\Theta-\pi^*\hat{\Theta})=\iota^*\Omega
-\pi^*\hat{\Omega}=0 \eeq
whence
\beq dS:=\iota^*\Theta-\pi^*\hat{\Theta} \eeq is (locally) is exact. S is the
Hamilton-Jacobi functional, it is the generator of a singular canonical
transformation from the large to the reduced phase space. Substituting
the momenta on $\Gamma$ by the the functional derivatives of S with respect
to the coordinates on $\Gamma$ solves the constraints because by doing
this substitution one pulled back the momenta to $\bar{\Gamma}$. Hence, one
way of obtaining the reduced phase space is to solve the Hamilton-Jacobi
equation for constrained systems.\\
Another method is suggested by looking at formula (3.36) : it says
that, up to a total differential, one obtains the reduced symplectic
potential simply by inserting the solution of the constraint equations
into the full symplectic potential. For field theories, there might also
be boundary terms involved in this reduction process, whose contribution to
the reduced symplectic structure does not vanish. They may be neglected
at a first stage because they will be recovered when one checks whether
the observables of the reduced phase space are finite and functionally
differentiable.

\subsection{The reduced phase space}

It will turn out that for this model the second method is more appropriate.
We are thus interested in the solutions of constraint equations.\\
We first take the following linear combinations of the vector and the
scalar constraint functional
\begin{eqnarray}
E^1 E^2 V+E^3 C & = & E(E^1 B^3+\frac{1}{2}E^3 B^1) \nonumber\\
-E^1 E^3 V+E^2 C & = & E(E^1 B^2+\frac{1}{2}E^2 B^1) \; ,
\end{eqnarray}
where $E=(E^2)^2+(E^3)^2$.\\
Setting these expressions strongly zero we obtain 2 possible solutions :\\
Case I : $E=0$ (degenerate case)\\
Looking at the formula for the metric (3.17) we see that there is no radial
distance now. From the reality of the triads we conclude further that
$E^2=E^3=0$ whence we conclude $E^1=E^1(t)$ via setting the Gauss constraint
equal to zero. Obviously this solution of the constraint equations is not
valid in the asymptotic ends since it violates the asymptotic conditions on
the fields. It can therefore only hold inside the hypersurface and we should
glue it to a solution of the constraints appropriate for the asymptotic
regions. For compact topologies it is a global solution of the constraints.
Applying the framework of the previous subsection we obtain for the reduced
symplectic potential
\beq \hat{\Theta}[\partial_t]=-iE^1\frac{d}{dt}\int_\Sigma dr A_1 \;.\eeq
Case II : $E\not =0$ (nondegenerate case)\\
We now conclude
\begin{eqnarray}
0 & = & E^1 B^3+\frac{1}{2}E^3 B^1 \nonumber\\
0 & = & E^1 B^2+\frac{1}{2}E^2 B^1
\end{eqnarray}
and can further distinguish between a) $B^1=0$ and b) $B^1\not=0$.\\
Subcase a)\\
Then either $B^2=B^3=0$ or $E^1=0$. Consider first the case $B^I=0$. Then
an elementary calculation shows (recall the abbreviations (3.16))
$(B^2)^2+(B^3)^2=(A')^2/A+A(A_1+\alpha')^2=2(A_1+\alpha')^2=0$ whence
$\gamma:=A_1+\alpha'=0$. Now, by writing the symplectic potential in terms
of 'cylindrical coordinates', that is, by plugging $(A_2,A_3)=\sqrt{A}
(\cos(\alpha),\sin(\alpha))\mbox{ and }(E^2,E^3)=\sqrt{E}
(\cos(\beta),\sin(\beta))$ into $\Theta=-i\int_\Sigma dr \dot{A}_I E^I$
we easily obtain up to a total differential
\beq \Theta=\int_\Sigma dr[\dot{\gamma}E^1+\dot{B}^1\sqrt{\frac{E}{A}}
\cos(\alpha-\beta)+\dot{\alpha}{\cal G}]\; . \eeq
Accordingly, the symplectic structure pulled back to the Gauss reduced phase
space vanishes identically if $B^1=0$. We are not interested in this trivial
case of a reduced phase space consisting of only one point any longer.\\
Subcase b)\\
We can divide by $B^1$ (everywhere except for isolated points) to solve eqs. 
(3.39) for the
momenta $E^2\;\mbox{and}\;E^3$
\beq
E^2=-\frac{2 E^1}{B^1} B^2\mbox{ and }
E^3=-\frac{2 E^1}{B^1} B^3
\eeq
and insert this into into the Gauss constraint :
\beq 0=B^1(E^1)'+2 E^1(B^1)' \; .\eeq
Eqn. (3.42) can be integrated :\\
\beq E^1=\frac{m^2}{(B^1)^2} \eeq
where the constant of integration m takes real values in the asymptotically
flat case for the following reason : we will show later that $B^1$ is real
on the constraint surface. Moreover, $E^1$ becomes $r^2$ at infinity.
Thus, $m^2$ must be a real positive constant.\\
The last step is then to pull back the symplectic potential. One can check
that modulo a total differential
\begin{eqnarray}
(\iota^*\Theta)[\partial_t] & = & -i m^2[\int_\Sigma dr[\dot{A}_1
\frac{1}{(B^1)^2}+\dot{A}_2\frac{-2B^2}{(B^1)^3}
+\dot{A}_3\frac{-2B^3}{(B^1)^3}] \nonumber\\
& = & -im^2\frac{d}{dt}\int_\Sigma dr\frac{\gamma}{(B^1)^2} \;.
\end{eqnarray}
One can check that the integrand in the last line of (3.44) vanishes
{\em on the contraint surface} as $1/r^2$ at infinity, and thus the integral
is well defined. However, it is functionally differentiable only if we
require $\delta a_2=\delta a_3=0$ in addition to the requirements derived
in section 3.1.

\subsection{Reality conditions}

The reality conditions in the degenerate case are obscure because the
spin connection coefficients are ill-defined (they are homogenous functions
of degree zero, so their value on the constraint surface depends on the
way the limit is taken). Therefore, we focus on the non-degenerate case in
the sequel.\\
First we prove that the magnetic fields are weakly real :\\
To begin with we have (recall (3.18))
\beq \bar{A}=(A_2-2\Gamma_2)^2+(A_3-2\Gamma_3)^2=A
+4[-A_2\Gamma_2-A_3\Gamma_3+(\Gamma_2)^2+(\Gamma_3)^2]=A+4\frac{(E^1)'}{E}
{\cal G} \eeq
which proves that $B^1$ is weakly real. We now solve the
constraints for the remaining magnetic fields
\[ B^2=-\frac{E^1}{2 E^2} B^1\mbox{ and }B^3=-\frac{E^1}{2 E^3} B^1 \]
to conclude from the reality of the triads that also $B^2, B^3$ are weakly
real.\\
We now exploit this result to show that $\gamma$ is weakly imaginary. Since
$A\gamma=A_2 B^2+A_3 B^3$ we have ($\approx$ means $=$ on the constraint
surface)
\beq \bar{\gamma}\approx-\gamma+2\frac{\Gamma_2 B^2+\Gamma_3 B^3}{A}
=-\gamma-2\frac{(E^1)'}{E A}V\;. \eeq
Accordingly, the momentum conjugate to $P:=m^2,\;Q:=-i\int_\Sigma dr
\frac{\gamma}{(B^1)^2}$, is weakly real. The reduced phase space is therefore
the cotangent bundle over the positive real line. However, it proves
convenient for the interpretation of our Dirac observables to make a
canonical transformation and to describe the reduced phase space in terms
of m and $T:=2mQ$.

\subsection{Interpretation}

Let us explore what the geometrical meaning of m is.\\
Using the fall-off properties of the fields of section 3.1 and in
particular eqn. 3.27 we find that $q_{rr}\rightarrow 1-\sqrt{2}a_3/r+
O(1/r^2)$. Comparing this with the asymptotical form of a Schwarzschild
metric of mass M we find
\beq a_3=-\sqrt{2}M \eeq
whereas from equation (3.43) we have that (recall that $E^1\rightarrow r^2
+O(r)$), $m^2=2(a_3)^2$, whence
\beq m=\pm 2M \eeq
i.e. m is twice the Schwarzschild mass of the given solution. If we apply
the positive censureship conjecture we find that m has range on half the
real line only (\cite{15}).\\
Next we have a look at the reduced action. Plugging the solution of the
constraint equations into the action and using the fall-off properties of
the fields we find the reduced Hamiltonian in the asymptotically flat
context to be equal to
\beq H=(N_\infty-N{-\infty})m  \eeq
where $N_{\pm\infty}$ is the lapse at $r=\pm\infty$. Accordingly, the
solution of the equations of motion for the canonical pair $(m,T)$ turns
out to to be
\beq m=\mbox{const. and }T=\mbox{const.}+\tau_+ - \tau_- \eeq
where $\dot{\tau}_\pm(t)=N_{\pm\infty}(t)$, that is, $\tau_\pm$ is the
eigentime of an asymptotic observer at positive or negative spatial infinity
(recall that $ds=\sqrt{-g_{tt}}dt$ is the eigentime interval associated with
the time label interval $dt$ and that $g_{tt}=-N^2$ at spatial infinity for
asymptotically vanishing shift). Thus, {\em on shell} T can be identified
with the difference $\tau$ of these eigentimes. In particular, T is a
constant
if and only if both clocks run at equal velocities, that is, $N_\infty
=N_{-\infty}$ as is the case for the Kruskal solution.\\
To summarize, spherically symmetric canonical gravity adopts the form of an
integrable system where the role of the action and angle variable
respectively is played by the mass and the difference of eigentimes
at both asymptotic ends respectively. In the case of closed topologies
the reduced Hamiltonian vanishes identically and T is a constant of the
motion, too.

\subsection{Quantization}

In the reduced phase space approach one quantizes the classically reduced
phase space. In our case, we have that the reduced phase space is a
cotangent bundle over either the real or half the real line. In the latter
case one would proceed to quantize the canonical pair $(\ln(\pm m),\pm m T)$
which again provides one with a cotangent bundle over the real line. So
there is nothing essentially new coming from this case and we therefore
concentrate on the first case.\\
We choose the representation in which $T$ is diagonal and arrive at
the following operator equivalents of our basic variables
\beq \hat{m}:=-i\hbar\frac{\partial}{\partial T},\; \hat{T}=T \;.\eeq
The physical Hilbert space consists of the usual complex-valued,
square integrable functions of T. The solutions of the Schroedinger
equation
\beq i\hbar\frac{\partial \Psi}{\partial t}(T)=\hat{H}\Psi(T)=
-i\hbar\frac{d\tau}{dt}\frac{\partial\Psi}{\partial T}(T)
\eeq
are given by
\beq \Psi(t,T)=f(T-\tau(t)) \eeq
i.e. it is an arbitrary function of the argument displayed. Of course,
only normalizable functions f should be considered. In particular
the eigenfunctions of the Hamiltonian, $f=\exp(ik(T-\tau))$ are not
normalizable.\\
\\
\\
In the operator constraint (Dirac) approach, one solves the constraints
after quantizing. Let us follow the steps of this quantization procedure.\\
Step1) : Quantize a complete set of basic operators such that its
commutator algebra mirrors the associated classical Poisson algebra. We
choose
\beq [\hat{A}_I(x),\hat{A}_J(y)]=[\hat{E}^I(x),\hat{E}^J(y)]=0,\;
[\hat{A}_I(x),\hat{E}^J(y)]=-\hbar\delta_I^J\delta(x,y)\;. \eeq
Step2) : Choose a representation of this algebra on a complex vector
space V. We choose the self-dual representation, that is, V consists of
holomorphic functionals of the connection. Our operators are then
represented as follows :
\beq (\hat{A}_I(x)\Psi)[A]:=A_I(x)\Psi[A],\;(\hat{E}^I(x)\Psi)[A]
:=\hbar\frac{\delta\Psi[A]}{\delta A_I(x)} \eeq
where $\delta/\delta A_I(x)$ is the functional derivative\footnote{If it
exists, the functional derivative is defined by $\int_\Sigma dr \xi(r)
\delta f[\phi]/\delta\phi(r):=\lim_{s\to 0}(f[\phi+s\xi]-f[\phi])/s$ for
any test function $\xi$ of compact support}.\\
Step3) : Try to find a consistent ordering of the constraints, that is,
they should form a commutator subalgebra in the sense that the constraint
operators appear always ordered to the right after commuting two constraints
(this is is a nontrivial requirement because the structure functions
for constraints bilinear in the momenta turn out to be operator valued,
compare Wipf's lectures). As we analyzed in section 3.3, for our model it
is actually possible to cast the constraints into a form in which they are
linear in the momenta. Therefore, we do not have any problems with this
step : just order the scalar constraint in such a way that the operators
linear in the momenta which are to vanish appear to the right handside and
order the momenta to the right in the remaining constraints. Thus, for
sector I we would write
\beq \hat{C}=\hat{E}^1(\hat{B}^2\hat{E}^2+\hat{B}^3\hat{E}^3)+\frac{1}{2}
\hat{B}^1(\hat{E}^2\hat{E}^2+\hat{E}^3\hat{E}^3) \eeq
whereas for sector II we would order as
\beq \hat{C}=\hat{E}^2(\hat{B}^2\hat{E}^1+\frac{1}{2}\hat{B}^1\hat{E}^2)
+\hat{E}^3(\hat{B}^3\hat{E}^1+\frac{1}{2}\hat{B}^1\hat{E}^3) \eeq
and one can explicitely check that the commutator algebra closes in the
sense that the operators linear in the momenta which are to annihilate
the physical states always appear to the right.\\
Note that we should actually regulate the scalar constraint since it is
bilinear in the momenta. However, since we are effectively working with
a rewritten version which is linear in the momenta, we can circumvent this
step.\\
Step4) : Solve the constraints, that is, find the physical subspace $V_{phys}$
of the vector space V.\\
For sector I this amounts to imposing
\beq \frac{\delta \Psi}{\delta A_2(x)}=\frac{\delta \Psi}{\delta A_3(x)}
=(\frac{\delta \Psi}{\delta A_1(x)})'=0 \eeq
the unique solution of which is given by
\beq \Psi[A]=f(\int_\Sigma dr A_1(r))\;. \eeq
For sector II the kernel of the constraint operators consists of the
functions satisfying
\begin{eqnarray}
0 & = & \hat{B}^2\frac{\delta\Psi}{\delta A_1}+\frac{1}{2}\hat{B}^1
\frac{\delta\Psi}{\delta A_2}\nonumber\\
0 & = & \hat{B}^3\frac{\delta\Psi}{\delta A_1}+\frac{1}{2}\hat{B}^1
\frac{\delta\Psi}{\delta A_3}\nonumber\\
0 & = & \hat{B}^1(\frac{\delta\Psi}{\delta A_1})'+2(\hat{B}^1)'
\frac{\delta\Psi}{\delta A_1}
\end{eqnarray}
the unique solution of which is given by
\beq \Psi[A]=f(\int_\Sigma dr \frac{A_1+(\arctan(\frac{A_3}{A_2}))'}
{(B^1)^2})\;. \eeq
Thus, for both cases the solution consists of arbitrary functions of the
functionals displayed.\\
Step5) : Find a complete algebra of basic quantum observables.\\
By definition, observables leave the physical subspace invariant.
Accordingly, we choose them to be the multiplication and differentiation
operators with respect to the argument of the functions of the physical
subspace. For sector I we thus have
\beq \hat{Q}:=-i\int_\Sigma dr A_1,\;\hat{P}:=\hat{E}^1(x) \eeq
whereas for sector II we obtain
\beq \hat{Q}:=-i\int_\Sigma dr\frac{A_1+
(\arctan(\frac{A_3}{A_2}))'}{(B^1)^2},\;\hat{P}:=(\hat{B}^1)^2(x)
\hat{E}^1(x)\;. \eeq
In both cases the argument x of the operator $\hat{P}$ is irrelevant since
it is a spatial constant on the physical subspace.\\
Step6) : Equip $V_{phys}$ with a Hilbert space structure by demanding that
the reality conditions {\em induced} on the quantum observables become
adjointness conditions with respect to that inner product.\\
It follows from the analysis in section 3.6 that for sector II the
classical analogues Q and P of the observables found in step 5 are
real. Since the observables $\hat{Q},\hat{P}$ found in step5 are canonically
conjugate, $[\hat{Q},\hat{P}]=i\hbar$, we will also postulate for sector I
that Q is classically real (its imaginary part arises then classically from
a canonical transformation). \\
The classical range of P is positive in case of sector II. Accordingly, we
either choose a representation in which $\hat{P}$ is diagonal and
proceed along the lines of \cite{8} or we allow for classically not allowed
regions of the quantum theory and can stay within the representation such
that $\hat{Q}$ is diagonal. Let us choose the latter option.\\
Then in both sectors, the unique inner product that accomplishes our aim
is just $L_2(R,dQ)$.\\
One could finally form the direct sum of both Hilbert spaces, thus producing
sectors in the technical sense of the word because states of different
sectors cannot be superposed (they are not annihilated by the same constraint
operators).\\
\\
Altogether, in the present model both quantization procedures give equivalent
answers.

\subsection{Loop representation}

In one dimension, except for the case of a closed topology, there are no
loops. However, in order to test the 3-dimensional theory, one should
rather look at loop variables for the 3+1 case restricted to spherical
symmetry. It turns out that one can find a subset of loops which form
a closed loop subgroup such that one can express all O(2)-invariant
quantities in terms of them. When expressing the Dirac observables found
for the present model in terms of them, the expressions become rather
horrible for the non-degenerate sector but become very simple for the
degenerate sector. This is in accordance to the fact that all solutions
to the constraints in the 3+1 case that have been found so far belong
to the degenerate sector but that no solutions are known for the
non-degenerate sector.\\
More details are given in \cite{10}. We refrain from giving them here
because issues like a loop transform and an algebra of loop operators have
not been worked out yet.

\subsection{Discussion}

The model of spherically symmetric gravity has been successfully quantized,
both, via the reduced phase space and the operator contraint approach.
The model captures various technical problems of the full 3+1 case :\\
Its reality structure is non-trivial, the constraints mirror those of the
full 3+1 theory, in particular they are bilinear in the momenta. What
comes out as a surprise is that the Dirac observables (3.63) have such a simple
reality structure. That raises the hope that also in the full theory the
reality structure of the reduced phase space turns out to be rather simple.\\
Also, all the results given here in terms of Ashtekar's variables can also be 
written in terms of geometrodynamical (ADM) variables. In particular, it
is possible to write down a one parameter family of exact solutions to the
Hamilton-Jacobi equation associated with the Wheeler-DeWitt equation 
(\cite{10}).\\
These are the positive remarks. Again sharp criticism is in order :\\
We actually exploited the fact that the scalar constraint could be cast
in such a form that it is linear in the momenta. This technical help will
not be available in the full theory and we cannot expect to solve the
quantization programme without regularizing and renormalizing the constraint
operators.\\
The other issue is of course that again we are effectively dealing with
quantum mechanics rather than quantum field theory.\\
\\
\\
\\
{\large Acknowledgements}\\
\\
It is my pleasure to thank the relativity group at the Max-Planck-Institute
for Astrophysics in Munich for having invited me to give a talk at the
117th WE-Heraeus seminar 'The Canonical Formalism in Classical and Quantum
General Relativity'. In particular, I would like to thank Dr.\ Helmut Friedrich
for his efforts in the organization of the meeting.\\
This work was supported by the Graduierten-Programm of the Deutsche 
Forschungsgemeinschaft (DFG).

\end{document}